\documentclass[preprint]{aastex}
\usepackage{amssymb,amsmath}
\usepackage{graphicx}
\usepackage{natbib}
\usepackage[OT1,T1]{fontenc}

\begin{document}

\title{Multi-wavelength Spectral Analysis of Ellerman Bombs Observed by FISS and IRIS}
\author{Jie Hong\altaffilmark{1,2}, M.~D. Ding\altaffilmark{1,2}, and Wenda Cao\altaffilmark{3}}
\affil{\altaffilmark{1}School of Astronomy and Space Science, Nanjing University, Nanjing 210023, China \email{dmd@nju.edu.cn}}
\affil{\altaffilmark{2}Key Laboratory for Modern Astronomy and Astrophysics (Nanjing University), Ministry of Education, Nanjing 210023, China}
\affil{\altaffilmark{3}Big Bear Solar Observatory, New Jersey Institute of Technology, 40386 North Shore Lane, Big Bear City, CA 92314-9672, USA}

\begin{abstract}
Ellerman bombs (EBs) are a kind of solar activities that is suggested to occur in the lower atmosphere. Recent observations using the Interface Region Imaging Spectrograph (IRIS) show connections of EBs and IRIS bombs (IBs), implying that EBs might be heated to a much higher temperature ($8\times10^{4}$ K) than previous results. Here we perform a spectral analysis of the EBs simultaneously observed by the Fast Imaging Solar Spectrograph (FISS) and IRIS. The observational results show clear evidence of heating in the lower atmosphere, indicated by the wing enhancement in H$\alpha$, Ca II 8542 \r{A} and Mg II triplet lines, and also by brightenings in the images of 1700 \r{A} and 2832 \r{A} ultraviolet continuum channels. Additionally, the Mg II triplet line intensity is correlated with that of H$\alpha$ when the EB occurs, indicating the possibility to use the triplet as an alternative way to identify EBs. However, we do not find any signal in IRIS hotter lines (C II and Si IV). For further analysis, we employ a two-cloud model to fit the two chromospheric lines (H$\alpha$ and Ca II 8542 \r{A}) simultaneously, and obtain a temperature enhancement of 2300 K for a strong EB. This temperature is among the highest of previous modeling results while still insufficient to produce IB signatures at ultraviolet wavelengths.
\end{abstract}


\section{Introduction}
Ellerman bombs (EBs) are a kind of short-lived small-scale solar activities that show intense brightenings in the H$\alpha$ wings \citep{1917ellerman}. Recent observations revealed that EBs are also visible in the wings of Ca II H/K \citep{2008matsumoto,2015rezaei}, Ca II 8542 \r{A} \citep{2013yang,2015kim,2015rezaei}, He I D$_{3}$ and He I 10830 \r{A} \citep{2016libbrecht}, and in the ultraviolet continuum of 1600 and 1700 \r{A} \citep{2013vissers,2015rezaei,2016tian}, while they are almost transparent in the optical continuum, neutral metal lines (Na I D and Mg I b; \citealt{2015rutten}), or hotter coronal lines (171 and 193 \r{A} channels; \citealt{2013vissers,2016tian}). The variety of visibility in different spectral lines suggest that EBs occur in the lower atmosphere, usually in the upper photosphere to the lower chromosphere. A review of studies on EBs can be found in \cite{2002georgoulis} and \cite{2013rutten}.

Most EBs appear near the polarity inversion line, and are associated with flux emergence \citep{2007pariat,2008watanabe,2013yang,2016reid,2016danilovic,2016yang} or moat flows \citep{2011watanabe,2013vissers}. Numerical simulations have shown that the newly emerged magnetic flux can interact with each other or with pre-existing magnetic fields, and cause magnetic reconnections \citep{2007isobe,2009archontis,2013nelson}. Line-of-sight mass flows as outflows from magnetic reconnections can be derived from spectral line profiles using the bisector method, which shows upflows in the lower chromosphere while downflows in the photosphere \citep{2008matsumoto,2013yang}. It is also found that EBs can be associated with a small-scale loop \citep{2015nelson}, a surge \citep{2008matsumoto,2013yang,2016pasechnik} or jets \citep{2011watanabe,2015reid}.

There are two approaches to estimate the temperature increase in EBs based on observed spectral lines. One is to calculate non-LTE semi-empirical models that tend to reproduce the observed line profiles as well as possible. Up to now, the available calculations indicated that the temperature increase is in the range of a few hundreds to 3000 K \citep{2006fang,2006socas,2013bello,2014berlicki,2015li,2016kondrashova}. The second approach is to fit the spectral lines using some simplified models. In this regard, \cite{2014hong} proposed the two-cloud model, which assumes that the EB atmosphere comprises two characteristic layers, a lower (relatively hot) layer responsible for the line wing emission and an upper (relatively cool) layer for the line center absorption. By fitting the observed H$\alpha$ line profiles, they derived a temperature increase in EBs (the lower layer) to be around 1000 K, consistent with previous results.

Recently, the Interface Region Imaging Spectrograph (IRIS; \citealt{2014depontieu}) discovered some ``hot explosions'' that are thought to occur in the lower atmosphere \citep{2014peter}. They are also referred to as IRIS bombs (IBs), with many similarities to EBs, except that the former exhibit enhanced UV line emissions. However, \cite{2015judge} suggested that IBs occur in the low-mid chromosphere or above, as a result of Alfv\'{e}nic turbulence. \citet{2015kim} studied an IB that appeared to be a very weak EB in the line wings of H$\alpha$. \citet{2015vissers} studied five EBs, all of which have corresponding IBs. They suggested that the top of the flame-like EBs can be heated to a high temperature so that they are also visible in IRIS Si IV lines. \citet{2016tian} studied ten IBs and found that six of them are associated with EBs. \cite{2016libbrecht} studied three EBs with IB signature and estimated the upper limit of temperature to be between $2\times10^{4}-10^{5}$ K by assuming that the line broadenings are caused by thermal Doppler motions. These observations suggest a possible link between EBs and IBs or, to say, that EBs might be hotter than what previous models predicted.

Therefore, two of the key questions regarding EBs include the forming height and temperature enhancement, which are associated with the mechanism of EBs as well as their relationship to IBs. We have investigated the temperature enhancement by fitting the H$\alpha$ profiles of EBs in our previous work \citep{2014hong}. However, there remains some uncertainty if using only one spectral line. In order to improve the reliability of the result, one needs to use more spectral lines with different formation heights. In this work, we fit the profiles of H$\alpha$ and Ca II 8542 \AA\ lines simultaneously. We also choose a very strong and a relatively weak one to see the difference in their spectral lines. More importantly, we need to check how much temperature enhancement the strong EB can generate.

For the above purpose, we investigate the EBs based on multi-lines observed simultaneously with the Fast Imaging Solar Spectrograph (FISS; \citealt{2013chae}) and IRIS. We present the imaging and spectral features of EBs in Section~2. In Section~3, we apply the two-cloud model to fit the H$\alpha$ and Ca II 8542 \r{A} lines simultaneously and present the results. Discussion on the implications of the results and a summary are given in Section~4.

\section{Observations}
The observations were performed simultaneously by FISS and IRIS, to a target active region NOAA 12401 (241E, 317S) on 2015 August 16. FISS is attached to the 1.6 m New Solar Telescope (NST; \citealt{2010cao,2012goode}) at Big Bear Solar Observatory, which can obtain two-dimensional spectral data from dual bands simultaneously using a scanning mode, resulting in high-resolution and high-cadence spectra of H$\alpha$ and Ca II 8542 \r{A}. The scanning has a cadence of 40 s, covering a field of view (FOV) of 40\arcsec$\times$40\arcsec, with a pixel size of 0\arcsec.16. The spectral resolution is 19 m\r{A} for H$\alpha$ and 26 m\r{A} for Ca II 8542 \r{A}.

IRIS observed the same region at that moment, and scanned over a small pore for one hour from 19:00 UT. IRIS can provide spectral scan and slit-jaw images (SJIs) at near-ultraviolet (NUV) and far-ultraviolet (FUV) lines simultaneously. These lines have formation temperatures of tens of thousands of Kelvins, and are thus good diagnostics of the upper chromosphere and transition region. In this observation, IRIS targeted a small sunspot using a 40 s cadence raster scan. The cadence of each scanning step is 1.2 s, which seems too short to ensure a sufficient signal-to-noise ratio for the lines in the FUV bands. In practice, we do a 3$\times$3 spatial bin for the FUV spectra, resulting in an actual spatial sampling of 1.05\arcsec$\times$1.05\arcsec. The NUV spectra and SJIs are not modified. The spectral scanning covers an FOV of 11\arcsec$\times$119\arcsec. All SJIs have a cadence of 10 s.

The Atmospheric Imaging Assembly (AIA; \citealt{2012lemen}) onboard the Solar Dynamics Observatory (SDO, \citealt{2012pesnell}) can provide extreme ultraviolet continuum images at 1700 \r{A}, with a time cadence of 24 s and a pixel size of 0\arcsec.6. Such EUV continuum is believed to originate from the temperature minimum region and upper photosphere \citep{2012lemen}. Besides, the Helioseismic and Magnetic Imager (HMI; \citealt{2012schou}) provides the line-of-sight magnetic field with the same pixel size but a cadence of 45 s.

Before further analysis, we need to co-align the images from different instruments using the same method as mentioned in \citet{2016hong}. In this case, the IRIS SJI 2832 \r{A} is used as the reference image. The accuracy of co-alignment between NST and IRIS images is within 0\arcsec.7.

\subsection{Imaging Observations}
\begin{figure}
    \centering
    \plotone{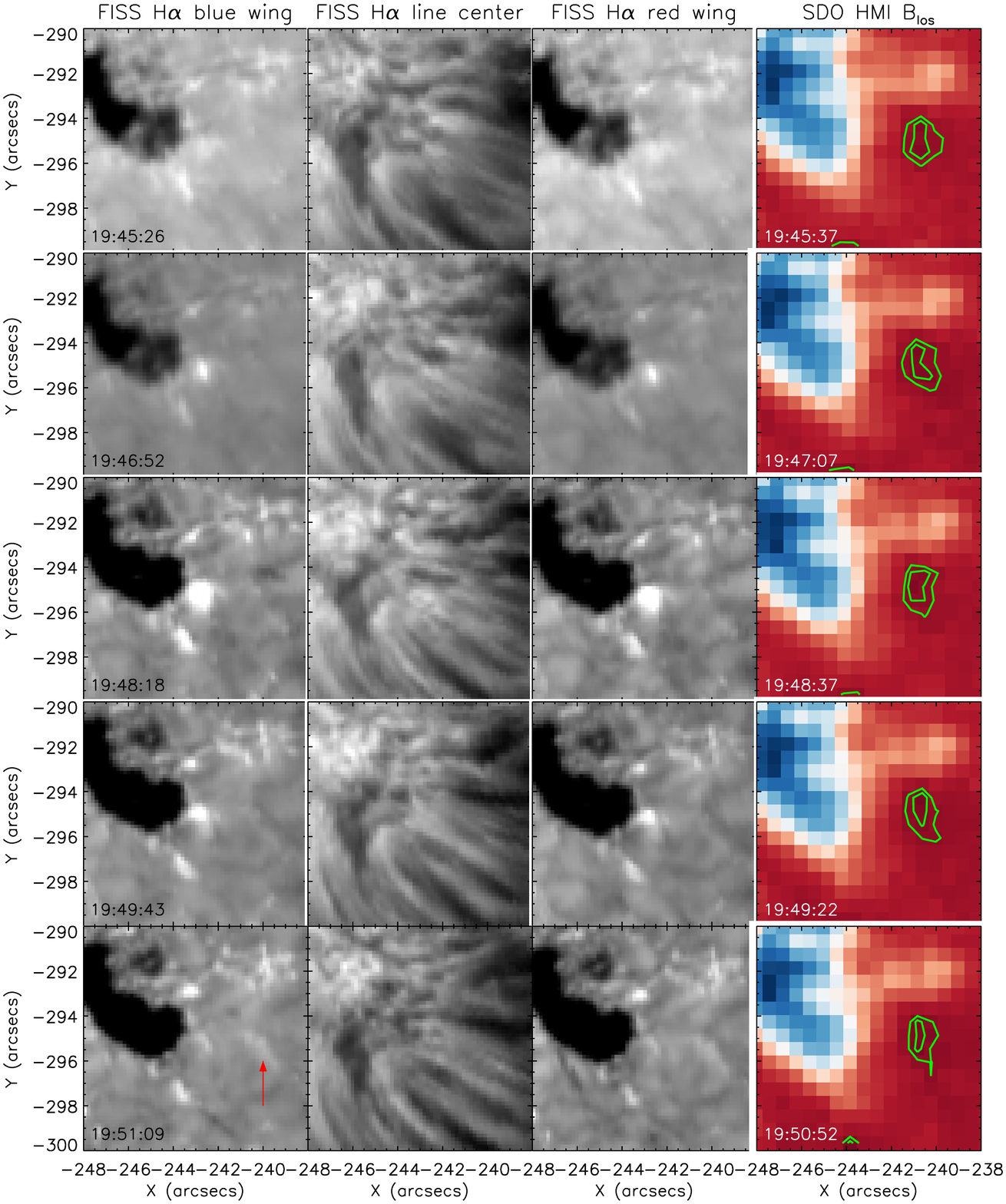}
    \caption{Time series of reconstructed images from FISS and HMI observations showing the evolution of the strong EB. From left to right, the four columns show H$\alpha$ images at the blue wing ($-1$ \r{A}), the line center, the red wing ($+1$ \r{A}) and line-of-sight magnetic field (red for negative and blue for positive polarity), respectively. The green lines are contours of magnetic field strengths of $-10$ and $-30$ G. The red arrow points to the weak EB that is marginally visible at $\pm1$ \r{A}.}\label{fig1}
\end{figure}

\begin{figure}
    \centering
    \plotone{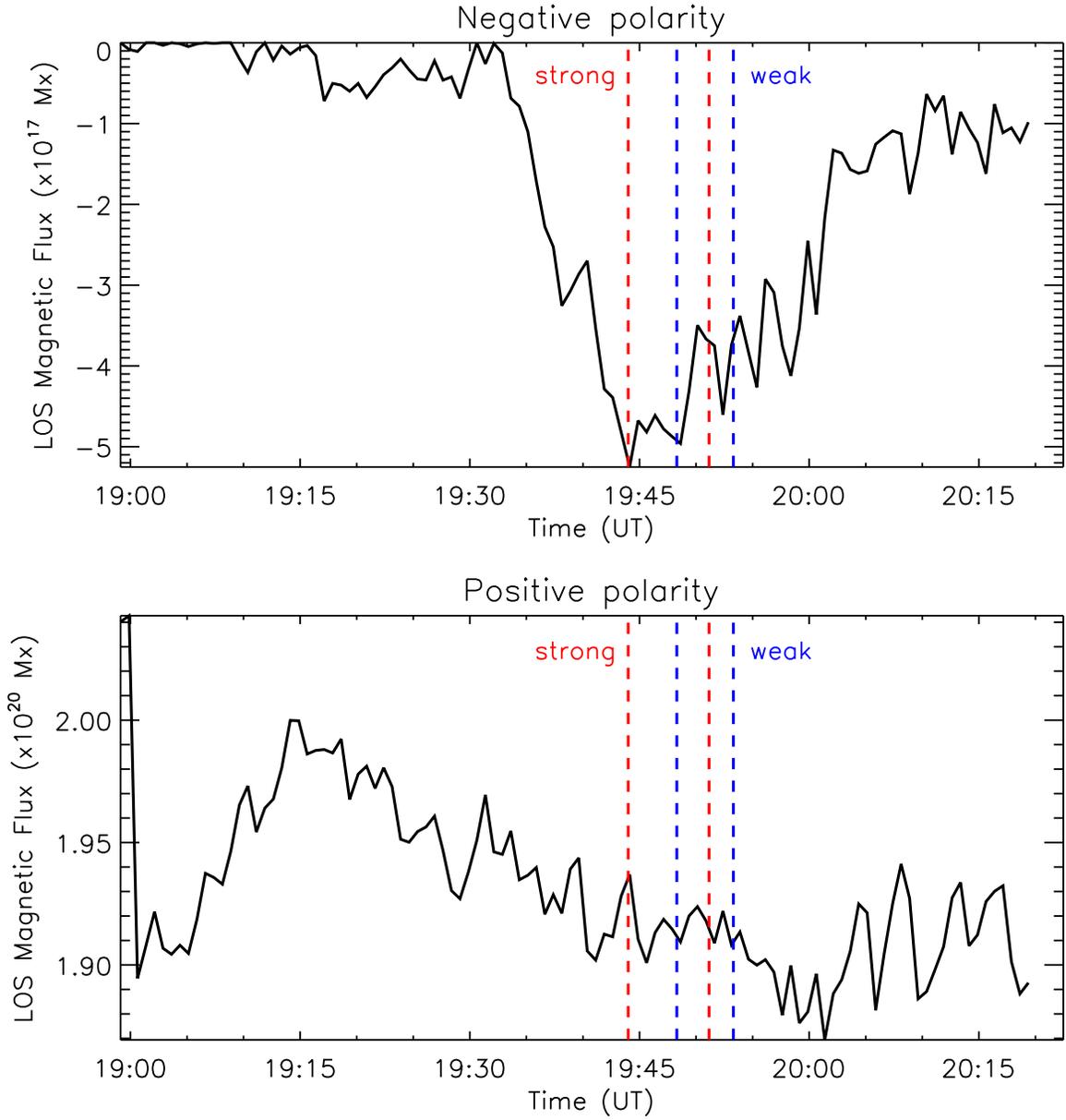}
    \caption{Time evolution of the magnetic flux of both the negative polarity and the positive polarity. The dashed lines in these two panels show the starting and ending time of the strong EB (red) and the weak EB (blue).}\label{fig1b}
\end{figure}

The active region we observed was quite productive, and had produced several small flares during the three days before  the observation time. Thus it is a promising region to detect Ellerman bombs. We identified one strong and one weak EB during the one-hour observation with their heliocentric angle to be about 30$^{\circ}$. Figure~\ref{fig1} shows the time evolution of the reconstructed images of the strong EB at different parts of the H$\alpha$ line, which reveal obvious brightenings at the line wings but no clear response at the line center, as well as the line-of-sight magnetic field. The lifetime of this EB is approximately six minutes, which is within the typical range. From the magnetic structure, we can find a small negative polarity near the pore, as delineated by the green contours. Compared to the strong magnetic field in the pore, the negative polarity is quite weak but still visible. Thus, the strong EB lies between two opposite polarities where magnetic reconnection could take place. The weak EB also lies near the rim of the negative polarity, but in the south of it (Figure~\ref{fig1}). The evolution of the line-of-sight magnetic flux of the negative and positive polarities is shown in Figure~\ref{fig1b}. The negative polarity newly emerges about 15 minutes before the moment when magnetic cancellation takes place and the strong EB occurs. The weak EB occurs later. The evolution of the positive polarity is not exactly the same as the negative polarity, because the positive flux mainly comes from the pore, which is relatively large in size and less influenced by the newly emerged negative polarity. However, the magnetic fluxes of both polarities are decreasing during the two EBs. One can also see from the contours in Figure~\ref{fig1} that the area of the negative polarity is gradually decreasing. The magnetic flux cancellation rate can be estimated to be about $2.2\times10^{14}$ Mx s$^{-1}$, within the typical range for EBs \citep{2016reid}.

\begin{figure}
    \centering
    \plotone{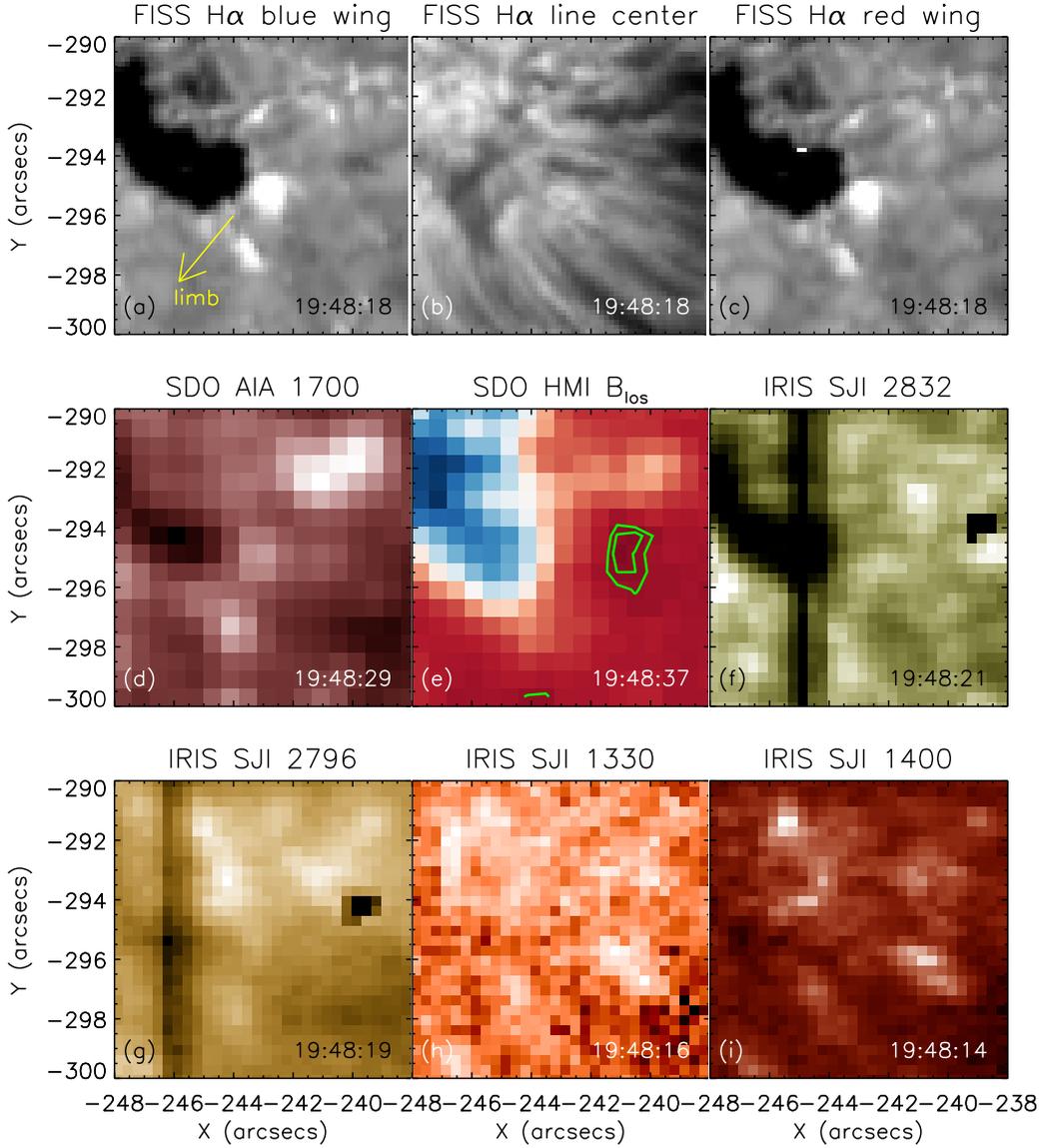}
    \caption{Images of the strong EB at its peak time observed by various instruments including FISS, AIA, and IRIS. The exact observing time of each instrument is depicted on the bottom right corner of each panel. (a-c) Reconstructed images at the H$\alpha$ blue wing, line center, and red wing. The wavelengths for the three FISS images are the same as in Fig.~\ref{fig1}. The yellow arrow in panel (a) points to the nearest solar limb. (d) AIA 1700 \r{A} image. (e) HMI line-of-sight magnetic field. Red is for negative polarity and blue for positive polarity. The green lines are contours of magnetic field strengths of $-10$ and $-30$ G. (f-i) IRIS SJI at 2832, 2796, 1330, and 1400 \r{A}.}\label{fig2}
\end{figure}

Figure~\ref{fig2} shows multi-wavelength images and the magnetic structure of the EB at the peak time of H$\alpha$ intensity. We also detect brightenings of the EB in the AIA 1700 \r{A} and IRIS SJI 2832 \r{A} images (Fig.~\ref{fig2}(d) and (f)). As the formation layer of ultraviolet continuum is roughly at the temperature minimum region, brightenings in these wavebands provide evidence of heating in the lower atmosphere. However, for the remaining three IRIS SJIs (Fig.~\ref{fig2}(g)--(i)), there appears almost no response in the EB region. This implies that the heating of this EB is not strong enough for the local temperature to reach the formation temperature of the Mg II line, which is approximately 10000 K.

\subsection{Spectral Observations}
\subsubsection{H$\alpha$ and Ca II 8542 \r{A} lines}
We show the FISS and IRIS line spectra for the strong EB in Figure~\ref{fig3}, and those for the weak EB in Figure~\ref{fig4}. The dashed lines in these two figures denote the pre-EB line profiles, which are the profiles at the same position but before the occurrence of EBs. It can be seen that for both EBs, the H$\alpha$ line wings are significantly enhanced at the peak time as compared to the pre-EB profiles, while the line center remains almost unchanged. Similar features appear in the Ca II 8542 \r{A} line, although the line wing enhancement is less significant than in the H$\alpha$ line. In both lines, the intensity enhancement extends over broad wings with, however, a gradually decreasing magnitude toward far wings, indicating that the heating occurs within a restricted layer in the lower atmosphere.

\subsubsection{Mg II lines}
\begin{figure}
    \centering
    \plotone{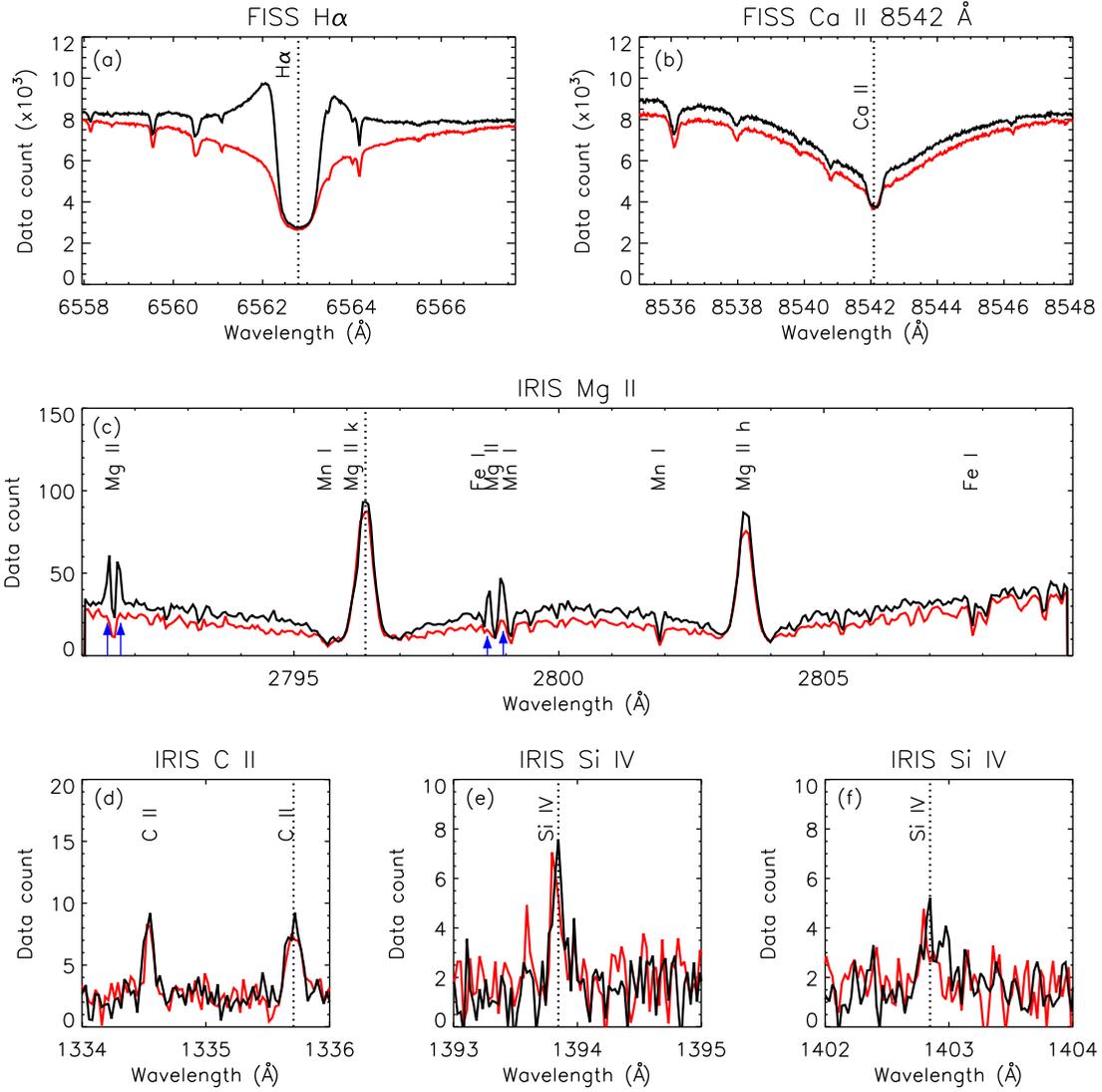}
    \caption{Spectra of the strong EB at its peak time from FISS (top row) and IRIS. Black curves refer to the EB spectra, while red curves are for the spectra at the same location before the EB (pre-EB profiles). The vertical dotted lines denote the line centers for some selected lines. The blue arrows in panel (c) indicate the enhancement in the Mg II triplet line wings.}\label{fig3}
\end{figure}

\begin{figure}
    \centering
    \plotone{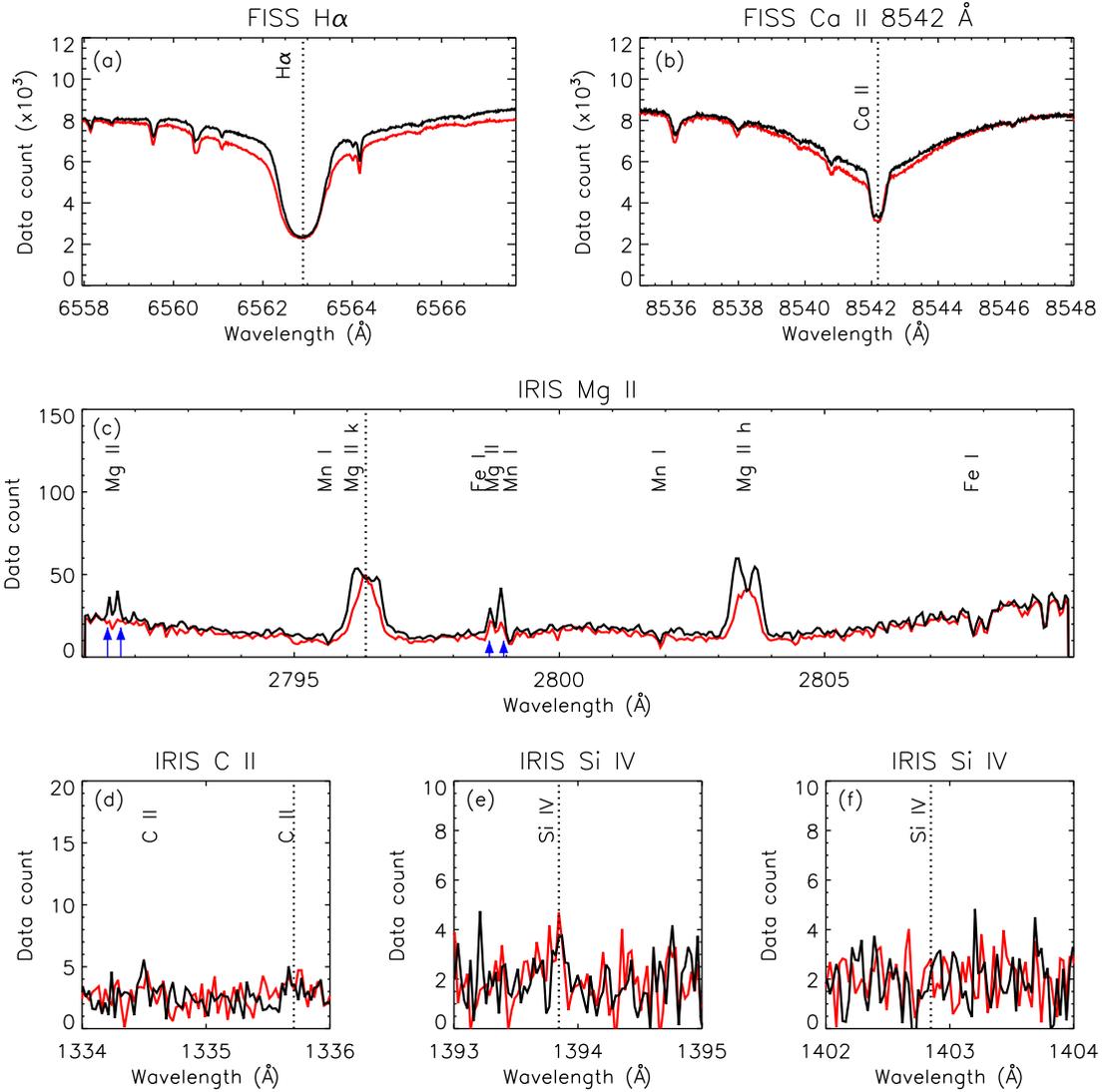}
    \caption{Same as Fig.~\ref{fig3} but for the weak EB.}\label{fig4}
\end{figure}

For the strong EB, spectra at the Mg II 2796 window (Fig.~\ref{fig3}(c)) show an enhancement of approximately 20\% in the ultraviolet continuum, corresponding to the brightening in the SJI 2832 \r{A}. However, the most striking feature in this wavelength window is the enhancement of the Mg II triplet lines. These lines are optically thick and form just above the temperature minimum region \citep{2015pereira}. \citet{2015pereira} found that for these lines, if there is heating in the lower atmosphere, there appears emission in the line wings while the line center remains in absorption, which is just what we observed. We find further that the triplet lines have a profile shape that resembles that of H$\alpha$, which has an obvious enhancement in the line wings but less influenced in the line center (denoted by the blue arrows in Fig.~\ref{fig3}(c)). By contrast, for the Mg II k/h lines, we can hardly see obvious change in the line intensity when the strong EB occurs. The width of the Mg II k/h lines does not change either during the strong EB. For the weak EB, besides the enhancement in the Mg II triplet line wings, there is also enhancement in the Mg II k/h line wings (Fig.~\ref{fig4}(c)).

\begin{figure}
    \centering
    \plotone{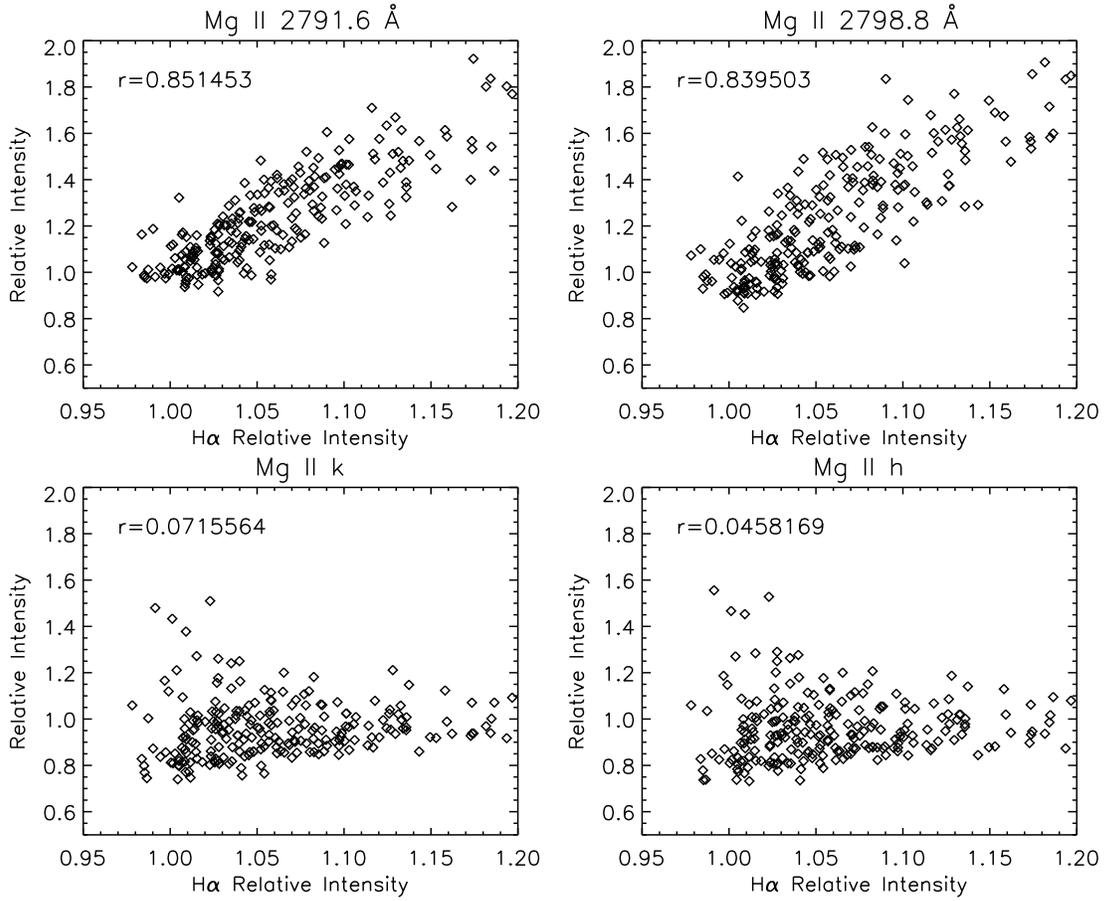}
    \caption{Scatter plot of the line intensities of Mg II triplet and k/h lines versus the H$\alpha$ line intensities in the EB region. All the intensities are integrated with wavelength and normalized to the pre-EB values. The Pearson correlation coefficient is also shown in each panel.}\label{fig5}
\end{figure}

Figure~\ref{fig5} shows a scatter plot of the wavelength-integrated intensities of the Mg II triplet lines versus that of the H$\alpha$ line in the EB region. The integration ranges for Mg II lines are carefully selected so that the influence of continnum is reduced to the least. In practice, we set the integration range to be $\pm0.25$ \r{A} for the Mg II triplet lines, and $\pm0.5$ \r{A} for the Mg II k/h lines. The intensities have been normalized to the corresponding pre-EB values. All the data points in Figure~\ref{fig5} come from the 16 pixels that cover the region of the strong EB over the whole lifetime. Clearly, there is a strong correlation between the Mg II triplet lines and the H$\alpha$ line. The few points with relative intensities less than 1.0 in the upper two panels comes mainly from the pixels that are located at the edge of the EB. In addition, there seems to be no correlation between the Mg II k/h lines and the H$\alpha$ line for the strong EB studied here.

\subsubsection{C II and Si IV lines}
As already reflected from the SJI 1330 and 1400 \r{A} images, the profiles of C II and Si IV lines suffer no change in the strong EB. Despite the large noise, these lines show no Doppler shift, no increase in line width or intensity (Fig.~\ref{fig3}(d--f)). This implies that the strong EB under study is not an IB that is heated to the transition region temperatures. The weak EB here is not an IB either (Fig.~\ref{fig4}(d--f)).

\section{Data Analysis}
\subsection{Two-cloud Model}
We use the two-cloud model proposed by \cite{2014hong} to investigate the temperature enhancement caused by the EB in the lower atmosphere. The lower cloud represents the energy release region while the upper cloud represents some overlying chromospheric fibrils. The temperature enhancement of the lower cloud can thus be deduced from the increase of the source function there, after fitting the observed EB profiles by synthetic profiles from the two-cloud model. The line profiles used in this model are contrast profiles, defined as
\begin{equation}
C(\lambda)=\frac{I_{EB}-I_{q}}{I_{q}},
\end{equation}
where $I_{EB}$ is the original profile at the brightest point in the EB region, and $I_{q}$ is the profile at the same location before the EB. As described in \cite{2014hong}, the physical parameters within each cloud are assumed to be constant. Note that \cite{2014hong} only fit the H$\alpha$ line. In this work, we make a significant improvement by fitting the H$\alpha$ and Ca II 8542 \r{A} lines simultaneously. Doing so can reduce the uncertainties in the fitting parameters, in particular, the temperature. Since the lower cloud is responsible for the origin of line wing emissions of both lines, the temperature increase there can be more constrained by observations considering that the two lines have different sensitivities to the local temperature.

As was done in \cite{2014hong}, we need to reduce the number of free parameters in the model in order for the fitting to be practical. The departure coefficients of the lower and upper levels of the hydrogen atom responsible for H$\alpha$ are obtained from the VAL-C model \citep{1981vernazza}. \citet{2016rutten} calculated the departure coefficients of the lower and upper levels of Ca II associated with the 8542 \r{A} line, and showed that the coefficients approach unity near the temperature minimum region where EBs are thought to occur (see the second row of their Fig.~1). For simplification, we assume that the departure coefficients of Ca II to be unity in the lower cloud.

As is well known, the emergent intensity relies on two key parameters, the line source function and the optical depth of the cloud. Sometimes, it is difficult to distinguish which one plays a major role in producing the intensity increase. This makes the fitting results not unique if the two parameters are both free. In fact, the line source function and the absorption coefficient depends on the local temperature in different manners. Here, we first check how the optical depth of the cloud may vary with the local temperature. Under the condition of local thermodynamic equilibrium (LTE), the line absorption coefficient (or optical depth if given a fixed geometric depth) should vary sensitively on the local temperature that is the sole factor determining the excitation and ionization of an atom, as clearly revealed in the calculations by \citet{2016rutten}. However, in non-LTE cases, radiative excitation and ionization may be important, which depends on the radiation field from beyond the local region. This makes the problem more complex.

Here, we make a test to show how the line absorption coefficient (optical depth) could vary under non-LTE circumstances. We solve the statistical equilibrium and charge conservation equations for the lower cloud simultaneously. We consider two atoms, hydrogen and singly ionized calcium. The number density of hydrogen is set to be $10^{15}$ cm$^{-3}$, which is a typical value in the temperature minimum region \citep{1981vernazza}. A model atom of three levels plus continuum for hydrogen and a model atom of five levels plus continuum for Ca II are adopted. Because of a quite low temperature in this region, metals have a larger contribution to electron density than hydrogen. Here, we assume an extra contribution of $10^{-4} N_{H}$ to the electron density from metals in the charge conservation equation. The photoionization cross sections are taken from \citet{1974shine}. All other atomic data are adopted from \citet{2000cox}. At the region near the temperature minimum, the background radiation field is mostly from the photosphere, while it has little dependence on the local temperature. Here, we assume a constant radiation temperature of 5700 K.

\begin{figure}
    \centering
    \plotone{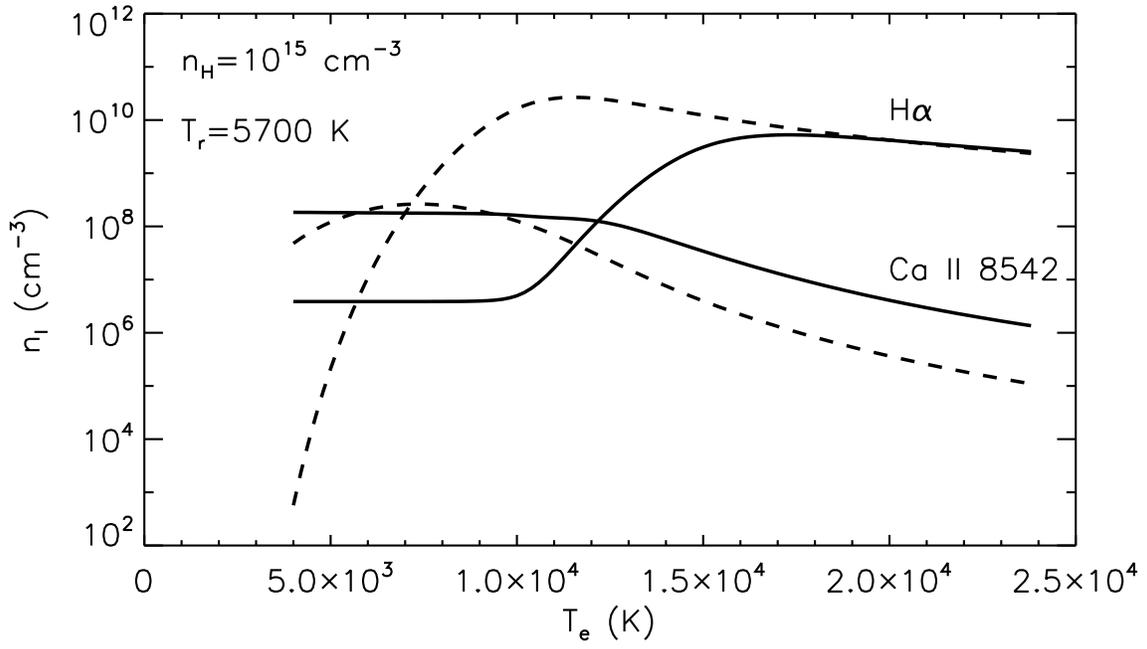}
    \caption{Number densities of hydrogen and singly ionized calcium at the lower level for the H$\alpha$ and Ca II 8542 \r{A} lines, respectively, as a function of the local electron temperature. Dashed lines refer to the results under LTE while solid lines are for the non-LTE results. In non-LTE cases, a constant radiation temperature (5700 K) is adopted. See text for details.}\label{fig6}
\end{figure}

Figure~\ref{fig6} shows the number densities of the lower levels associated with the spectral lines of interest in dependence on the local temperature. It is seen that under LTE, the curves (dashed) are similar to previous results (also see Fig.~5 of \citet{2016rutten}), while in non-LTE cases, the curves (solid) are quite different. Qualitatively, in the lower temperature domain ($T<10000$ K), the number density at the lower level varies insensitively to the local temperature under non-LTE, in sharp contrast to the LTE case. This is to say, even there occurs a temperature increase of several thousand Kelvins in the EB, there seems very little increase in the number density of the lower level. We then calculate the value of the optical depth at line center for the lower cloud, assuming the geometric thickness of the EB region to be 100 km near the temperature minumum region \citep{2016rutten}. The result is 0.5 for H$\alpha$ and 2.2 for Ca II 8542 \r{A}, which are then adopted as the optical depth of the lower cloud for the pre-EB atmosphere. The optical depth of the lower cloud for the EB atmosphere is a varying parameter that can be fitted reasonably as long as it does not vary sharply as shown above.

\subsection{Fitting Results}
\begin{figure}
    \centering
    \plotone{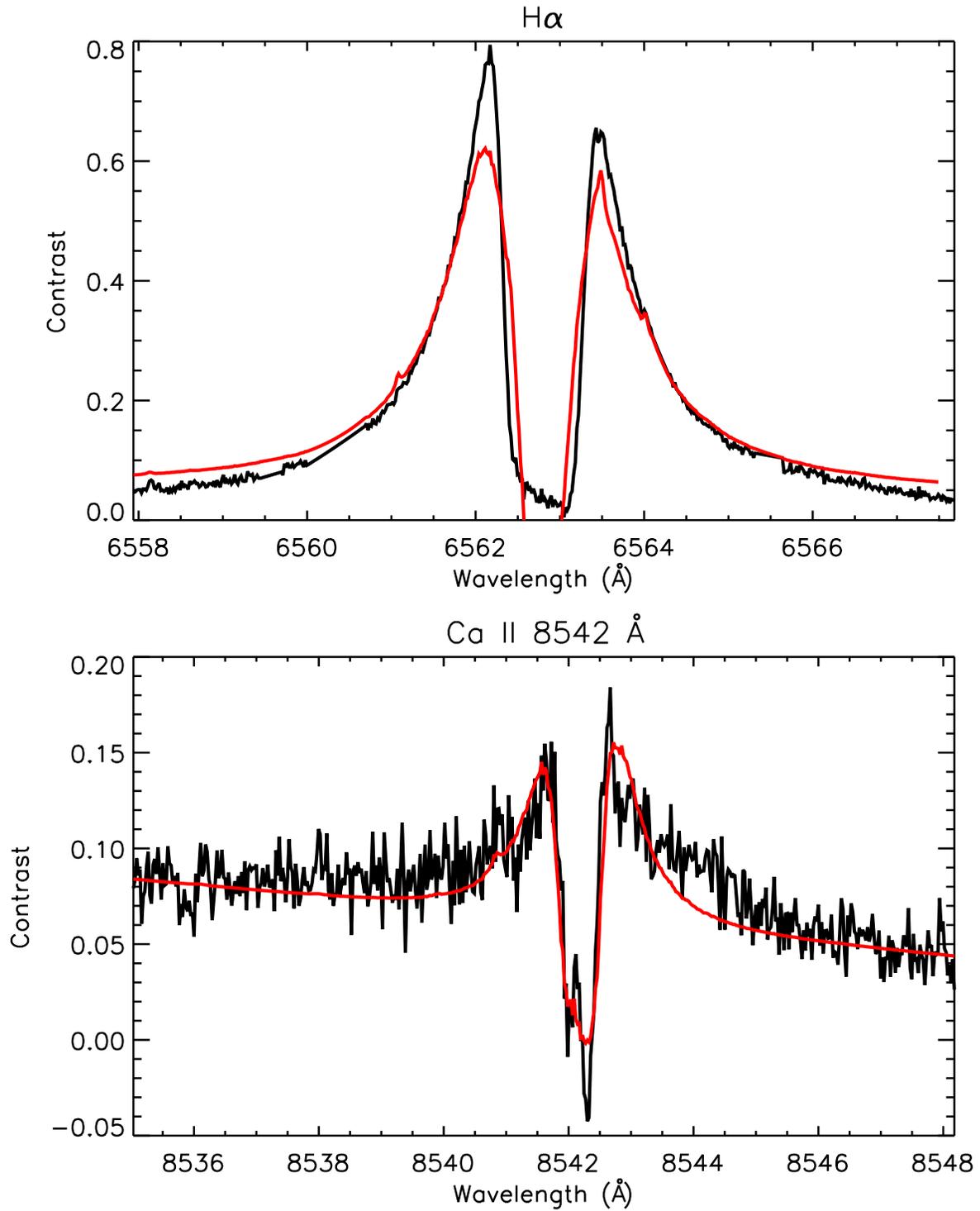}
    \caption{Observed H$\alpha$ and Ca II 8542 \r{A} contrast profiles (black) in the strong EB (at its peak time) and their fitting results (red) using the two-cloud model.}\label{fig7}
\end{figure}

The two chromospheric lines and their fitting results for the strong EB at its peak time are shown in Figure~\ref{fig7}. As was done in \cite{2014hong}, we introduce random noises to the profiles and repeat the fitting 10000 times. This is used to derive the standard deviation of the fitting results. Based on the parameters for the best fitting, we deduce a temperature enhancement of 2300$\pm20$ K in the lower cloud where the EB occurs. Such a temperature increase over the quiescent status is quite large but still within the typical range. We also find that the optical depth of the two chromospheric lines vary differently. For H$\alpha$, the optical depth at line center rises from 0.5 to 0.7$\pm0.05$, while for Ca II 8542 \r{A}, it decreases from 2.2 to 0.4$\pm0.07$. This can be explained by the change of the lower level populations. When there is a temperature increase, more hydrogen atoms will be exited to the lower level of H$\alpha$ from the ground level, while ionization is dominant for Ca II, resulting in a relatively smaller population at the lower level of Ca II 8542 \r{A}.

We should mention that the fitting results are more or less influenced by the hydrogen number density and the radiation temperature, because these two parameters are essential to the optical depth. We have also tested different values of them. Generally speaking, a larger temperature increase is expected when the initial optical depth is smaller. Quantitatively, the deduced temperature increase can change by about 500 K when the hydrogen number density is changed by three times or the radiation temperature is changed by 5\%. We think that near the temperature minimum region of the EB, it is unlikely that these two parameters suffer much more changes.

\section{Summary and Discussions}
We analysed the spectral data of EBs observed by NST and IRIS and fit the two chromospheric lines simultaneously using the two-cloud model. The brightenings in AIA 1700 \r{A} and IRIS SJI 2832 \r{A} images and the enhancement in the line wings of Mg II triplet lines confirm that there is heating in the lower atmosphere. For the strong EB studied, the relative enhancement at the near wing ($\sim\pm0.8$ \r{A}) of H$\alpha$ can reach as large as 0.8, one of the largest values among reported observations. However, we do not find clear response of IRIS Mg II k/h and hotter lines in our case, except for some enhancement in the wings of Mg II triplet lines. The spectral fitting with the two-cloud model yields a temperature enhancement of 2300 K in the temperature minimum region, still far below the formation temperature of the Mg II k/h or Si IV lines.

The enhancement of Mg II triplet lines in the EB region is consistent with previous observation and simulation results \citep{2015vissers,2015pereira}. The formation height of these triplet lines lies in the lower atmosphere \citep{2015pereira}, similar to the formation height of H$\alpha$ line wings. In fact, we find that in the EB region, the intensity of Mg II triplet lines correlates well with that of H$\alpha$. This suggests an alternative way to identify EBs using the Mg II triplet that can be routinely observed by IRIS. A statistical study based on more events is required to validate this proposition.

Recently, there is a heated discussion on the relationship between EBs and IBs. In some events, EBs and IBs are indeed closely related or refer to the same phenomenon. Some of the EBs studied by \citet{2015vissers} showed typical IB features, with a large width of Si IV profiles and an enhancement in Mg II k/h wings. However, although the strong EB studied here seems to have a large temperature enhancement, there appear almost no IB features. In fact, \citet{2016tian} found that only a small fraction of EBs are heated to IB temperatures, while most EBs do not have any IB signature. The strong EB studied here just belongs to the latter category, although a fairly large temperature increase is derived by spectral fitting.

It is interesting that, however, for the weak EB, we do see some weak enhancement in the Mg II k/h line wings. Considering that the H$\alpha$ and Ca II 8542 \r{A} lines are weaker in this EB, the derived temperature increase should be lower than in the strong EB. Therefore, our model, simplified by composition of two clouds but essentially one-dimensional, cannot explain the observed broader Mg II h/k line profiles without worsening the fitting to the H$\alpha$ and Ca II 8542 \AA\ lines.

In the future, it should be an interesting topic to further clarify the relationship between EBs and IBs. In this regard, more coordinated observations by ground-based telescopes and IRIS are needed. In particular, for the IBs, a synthetic model is expected to combine both heating in the chromospheric lines and possible heating that produces IRIS line emissions. Radiative hydrodynamic simulations are also required to show how the two different temperature structures are physically coupled. In particular, in order to simultaneously explain all the observed lines, one probably needs to take into account the inhomogeneity and multidimensionality of the atmosphere related to EBs.

\acknowledgments
We are very grateful to the referee for valuable comments. The authors thank the BBSO staff for their help during the observations. The BBSO operation is supported by NJIT and US NSF AGS-1250818 grants, and the NST operation is partly supported by the Korea Astronomy and Space Science Institute and Seoul National University and by the strategic priority research program of CAS with Grant No. XDB09000000. IRIS is a NASA small explorer mission developed and operated by LMSAL with mission operations executed at NASA Ames Research center and major contributions to downlink communications funded by ESA and the Norwegian Space Centre. SDO is a mission of NASA's Living With a Star Program. This work was also supported by NSFC under grants 11373023, 11403011, and 11533005, and NKBRSF under grant 2014CB744203.

\clearpage

\end{document}